\documentclass{kapproc}
\kluwerbib

\begin{document}

\input epsf

\articletitle{Galactic Magnetic Fields \ \\ as a consequence of Inflation}
%\articlesubtitle{This is an Article Subtitle}

\author{Konstantinos Dimopoulos}
\affil{Astroparticle \& High Energy Physics Group\\
Instituto de Fisica Corpuscular, Universitat de Valencia/CSIC\\
Edificio Institutos de Paterna, Apdo 22085, 46071 Valencia, 
Spain\footnote{This work was supported by 
the DGICYT grant PB98-0693 and by the E.U. under the contract HPRN-CT00-00148.
}}
\email{kostas@flamenco.ific.uv.es}

%\author{Anne-Christine Davis}
%\affil{Department of Applied Mathematics and Theoretical Physics\\
%Centre for Mathematical Sciences, University of Cambridge\\
%Wilberforce Rd, Cambridge CB3 0WA, England U.K.}
%\email{acd@damtp.cam.ac.uk}

%\author{Tomislav Prokopec}
%\affil{Institut f\"{u}r Theoretische Physik,
%Ruprecht-Karls-Universit\"{a}t Heidelberg\\
%Philosophenweg 16, D-69120 Heidelberg, Germany}
%\email{T.Prokopec@thphys.uni-heidelberg.de}

%\author{Ola T\"{o}rnkvist}
%\affil{Theoretical Physics Group\\
%Imperial College of Science, Technology and Medicine, University of London\\
%Prince Consort Rd, London SW7 2BZ, England U.K.}
%\email{o.tornkvist@ic.ac.uk}

\begin{abstract}
The generation of a magnetic field in the Early Universe is considered,
due to the gravitational production of the Z-boson field during 
inflation. Scaled to the epoch of galaxy formation this 
magnetic field suffices to trigger the galactic dynamo and explain the 
observed galactic magnetic fields. The mechanism is independent
of the inflationary model. 
\end{abstract}


\medskip

\section{Introduction}

Magnetic fields permeate most astrophysical systems \mbox{(\cite{kron})}. 
In particular galaxies carry magnetic fields of the order of $\sim\mu$Gauss
\mbox{(\cite{beck})}. In spirals such fields follow closely the density waves, 
which strongly suggests that galactic magnetic fields are sustained by a 
dynamo mechanism \mbox{(\cite{dynamo})}. The galactic dynamo combines the 
turbulent motion of ionized gas with the differential galactic rotation to 
amplify exponentially a week seed field up to dynamical equipartition value. 
This seed field should be coherent over the dimensions of the largest turbulent 
eddy ($\sim$100 pc) or else it may destabilize the dynamo action 
\mbox{(\cite{kuls})}. Moreover, in order to produce the observed galactic 
fields the seed field has to be stronger that a critical value. For a 
spatially-flat, dark-energy dominated Universe the
required strength may be as low as \mbox{$B_{\rm seed}\sim 10^{-30}$Gauss} 
(\cite{dlt}). However, the origin of such a field remains elusive.

Attempts to generate $B_{\rm seed}$ astrophysically via vorticity 
\mbox{(\cite{rees})} or battery \mbox{(\cite{batt1}, \cite{batt2})} 
effects require large-scale separation of charges i.e. substantial ionization, 
which is hard to realize as late as galaxy formation. Thus, the origin of the 
seed field is most likely primordial. Since it breaks isotropy the 
generation of large-scale magnetic fields has to occur out of thermal 
equilibrium. Therefore, before decoupling, magnetogenesis is possible either
{\em at phase transitions} or {\em during inflation} (for a review see 
\cite{rep}). 

Because phase transitions occur very early in the Universe history the 
comoving size of the particle Horizon is rather small. Thus, since 
magnetogenesis mechanisms are causal, the resulting magnetic field is too 
incoherent. On the other hand, inflation is possibly the only way one can 
achieve superhorizon correlations. However, the conformal invariance of 
electromagnetism forces the magnetic field to satisfy flux conservation 
{\em during} inflation (\cite{TW}). As a result the strength of the generated 
magnetic field is exponentially suppressed due to the rapid inflationary 
expansion of the Universe. 

We show that natural magnetogenesis during inflation can occur 
due to the breaking of the conformal invariance of the $Z$-boson field of the
Standard Model, which also contributes to the formation of large-scale
magnetic fields. The resulting seed field is sufficient to trigger the 
galactic dynamo and explain the observations in a model-independent way. 

We will use a negative signature metric and units such that 
\mbox{$c=\hbar=1$}.
%and \mbox{$G=6.72\times 10^{-39}$GeV$^{-2}$}.

\section{Inflation}

Inflationary theory is so successful that it may be considered as part of the 
standard model of Cosmology. Indeed, inflation manages with a single stroke
to solve the Horizon, Flatness and Monopole problems of the Standard Hot Big 
Bang (SHBB), while successfully providing the seeds for the formation of Large 
Scale Structure (i.e. the distribution of galactic clusters and superclusters) 
and for the observed anisotropies of the Cosmic Microwave Background 
Radiation (CMBR).

\paragraph{The basic picture}

According to inflationary theory, at some time in the early stages of its 
evolution, the Universe was dominated
by false-vacuum energy density, which played the role of an effective 
cosmological constant leading to a period of superluminal accelerated
expansion. 

In the simplest case one can consider that, during inflation, the energy 
density is 
\mbox{$\rho_{\rm inf}=\Lambda_{\mbox{\footnotesize eff}}/8\pi G=$ constant}, 
which suggests, by means of the Friedman equation:
\mbox{$H^2=(8\pi G/3)\rho_{\rm inf}$}, that the Hubble parameter 
\mbox{$H\equiv\dot{a}/a$} is constant and, therefore, the scale factor is 
\mbox{$a\propto\exp(Ht)$}, i.e. the Universe engages into a de-Sitter 
exponential expansion phase. 

During the inflationary expansion any pre-existing thermal bath is drastically
diluted and the temperature of the Universe is 
\mbox{$T\propto a^{-1}\rightarrow 0$}, i.e. the Universe is supercooled.
%At the end of the slow-roll period inflation is terminated and 
%the inflaton undergoes rapid coherent 
%oscillations, which are equivalent to massive particles (inflatons), that 
%decay
%and thermalize , creating thereby the thermal bath of the SHBB. 
When inflation ends there is enormous entropy production
and almost all the false vacuum energy is given to a thermal bath of newly 
created particles via a process called reheating. In typical inflationary 
models (e.g. chaotic, hybrid, natural) the reheating temperature is, 
\mbox{$T_{\rm reh}\sim \rho_{\rm inf}^{1/4}\sim 10^{16}$GeV}, i.e. at grand 
unification scale. 
%After the end of inflation the SHBB begins.

\paragraph{Magnetic Fields in Inflation}

Magnetogenesis during inflation is based on the fact that
the inflated quantum fluctuations of gauge fields become classical,
long-range gauge fields with {\em superhorizon} correlations. 
Unfortunately, this is not effective for conformally invariant gauge 
fields, such as the photon, since they do not couple to the inflating 
gravitational background. However, this is not so for the $Z$-boson of 
the Standard Model (SM), which may also contribute into magnetic field 
generation.

Indeed, due to supercooling, the electroweak (EW) symmetry is broken during 
inflation and, therefore, the $Z$-boson field is massive. The existence of
a non-zero mass $M_Z$ breaks conformal invariance and, consequently, $Z$ is 
gravitationally generated on superhorizon scales. However, at the end of 
inflation reheating typically restores the EW-symmetry 
(\mbox{$T_{\rm reh}>$ 100 GeV}) and the $Z$-boson is projected 
onto the Hypercharge giving rise to a hypermagnetic field
with superhorizon correlations. After the EW-transition the latter becomes a 
regular magnetic field.

\section{\boldmath $Z$-boson production in Inflation}

\paragraph{Initial amplitude at Horizon crossing}

During inflation all fields with masses smaller than $H$
are gravitationally produced (unless they are conformally invariant) because 
their Compton wavelength is larger than the Horizon size $\sim\!H^{-1}$ and,
therefore, their quantum fluctuations can reach the Horizon before 
dying out, i.e. the uncertainty principle allows the existence
of virtual particles long enough for them to exit the Horizon.
After their exit, the fluctuations cease to be causally 
self-correlated and cannot collapse back into the vacuum, that is, they become
from virtual, real classical objects. The energy of such particle generation 
is provided by the false vacuum energy driving inflation. 

Consider such a $Z$-boson fluctuation. Since the fluctuation is 
quantum-generated it is subject to the uncertainty relation,

\begin{equation}
\Delta{\cal E}\cdot\Delta t\simeq 1
\end{equation}
Because, typically, \mbox{$M_Z\ll H$} the fluctuation is not suppressed before 
reaching the Horizon and, also, the field can be considered to be effectively 
massless. Thus, the energy density of its fluctuation is mainly 
kinetic, \mbox{$\Delta{\cal E}\sim [\partial_t(\delta Z)]^2\Delta V$}.
Now, at Horizon crossing \mbox{$\Delta V\sim H^{-3}$}. Also, the time required 
for the fluctuation to reach and exit the Horizon is 
\mbox{$\Delta t\sim H^{-1}$}. Finally, because of the random nature of
quantum fluctuations (there is no coherent motion) we may identify 
\mbox{$\partial_t\sim\Delta t^{-1}$}. The above suggest, that 
\mbox{$(\delta Z)_H\sim H$}. 

In fact the amplitude of the fluctuation when exiting the Horizon is set by 
the Gibbons-Hawking temperature, \mbox{$T_H\simeq H/2\pi$}. This 
can be understood if the particle Horizon during inflation is viewed as the 
event horizon of an inverted (i.e. inside-out) black hole, in the sense that 
nothing can escape being ``sucked'' {\em out}. The above suggest,

\begin{equation}
|Z(t_{\rm x})|=(\delta Z)_H\simeq H/2\pi
\hspace{0.5cm}\mbox{and}\hspace{0.5cm}
|\dot{Z}(t_{\rm x})|=(\delta Z)_H/\Delta t\simeq H^2/2\pi
\label{intl}
\end{equation}
where $t_{\rm x}$ is the moment of Horizon crossing.
The amplitude of the fluctuation at Horizon crossing
is independent of $t_{\rm x}$ due to the self-similarity of de-Sitter 
spacetime (all dynamical scales, such as $H$, stay constant).

\paragraph{Superhorizon evolution during Inflation}

The subsequent evolution of the $Z$-fluctuation during inflation, after 
Horizon crossing, is classical and described by the equation of motion,

\begin{equation}
[\partial_\mu+(\partial_\mu\ln\mbox{$\sqrt{-D_g}$})]
[g^{\mu\rho}g^{\nu\sigma}(\partial_\rho Z_\sigma-\partial_\sigma Z_\rho)]
+M_Z^2g^{\mu\nu}Z_\mu=0
\end{equation}
where \mbox{$D_g\!\equiv$det($g_{\mu\nu}$)}. Using a Friedman-Robertson-Walker 
metric we get,

\begin{equation}
\partial_t^2Z_i-\partial_t\partial_iZ_t+H(\partial_tZ_i-\partial_iZ_t)+
a^{-2}(\partial_j\partial_jZ_i-\partial_i\partial_jZ_j)+M_Z^2Z_i=0
\label{eqm}
\end{equation}
Since the fluctuation in question is quantum-generated inside the Horizon 
it is causally connected at birth. Therefore, it can be taken to be smooth and 
homogeneous, when exiting the Horizon. Its subsequent, superhorizon evolution 
should not affect its comoving spatial distribution because of the symmetries
of the metric (Remember that, after exiting the Horizon, the fluctuation 
becomes causally disconnected). Thus, the initial homogeneity is expected to 
be preserved during the superhorizon evolution. So, we can take 
\mbox{$\partial_iZ_\mu=0$} which recasts (\ref{eqm}) as,

\begin{equation}
\ddot{Z}+H\dot{Z}+M_Z^2Z=0
\end{equation}
Solving this with \mbox{$M_Z\simeq$ const.} and the initial conditions  
of (\ref{intl}) we find, 

\begin{equation}
Z(t) = -\frac{H}{4\pi\nu}\Big(\frac{1}{2}-\nu\Big)
\,e^{-H\Delta t(\frac{1}{2}-\nu)}\;+\;\;
\frac{H}{4\pi\nu}\Big(\frac{1}{2}+\nu\Big)
\,e^{-H\Delta t(\frac{1}{2}+\nu)}
\label{Zt}
\end{equation}
where, \mbox{$\Delta t=t-t_{\rm x}$} with \mbox{$t_{\rm x}\leq t_{\rm end}$} 
and \mbox{$\nu\equiv\sqrt{\frac{1}{4}-(M_A/H)^2}$}.

The {\em physical} momentum-scale $k$ of fluctuation in question behaves as
\mbox{$k(t)\propto a^{-1}$}. Thus, because \mbox{$k(t_{\rm x})=2H$} and 
\mbox{$a\propto e^{Ht}$} we have, 

\begin{equation}
k(t)=\frac{a(t_{\rm x})}{a(t)}\,2H\Rightarrow e^{-H\Delta t}=\frac{k(t)}{2H}
\label{kHeDt}
\end{equation}
Inserting this into (\ref{Zt}) and considering \mbox{$M_Z\ll H$} we find,

\begin{equation}
Z(k)=-\frac{H}{2\pi}\Big(\frac{M_Z}{H}\Big)^2
\,\Big(\frac{k}{2H}\Big)^{(M_Z/H)^2}+
\frac{H}{2\pi}
\Big(\frac{k}{2H}\Big)
\label{Zk}
\end{equation}

\paragraph{\boldmath The photon versus the $Z$-boson}

During inflation $M_Z$ is not the bare mass of the $Z$-boson but it is given
by the magnitude of the EW-Higgs field condensate, 
\mbox{$M_Z\!=\!g_{\rm z}\sqrt{\langle\Psi^\dag\Psi\rangle}$}, 
where \mbox{$g_{\rm z}\!\sim\!0.6$} is the
gauge coupling of the $Z$ with the EW-Higgs field $\Psi$. Because $\Psi$ is
also effectively massless during inflation, every e-folding 
(= exponential expansion) creates a superhorizon fluctuation of order
\mbox{$H/2\pi$}, i.e. much larger than the {\small VEV} of the EW-Higgs field. 
The quantity $\!\sqrt{\langle\Psi^\dag\Psi\rangle}$ represents
an accumulative ``memory'' of these fluctuations corresponding to a 
random walk 
in the inner-space of $\Psi$ with number of steps given by the  
elapsing e-foldings, $\sim\!\ln[a(t_{\rm x})/a(t_{\rm i})]$, where 
$t_{\rm i}$ denotes the onset of inflation. Thus,

\begin{equation}
(M_Z/H)^2=(\frac{g_{\rm z}}{2\pi})^2\ln(k_{\rm i}/k)
\end{equation}
For the scales of interest \mbox{$(M_Z/H)^2\sim 0.05\gg (k/2H)$} and 
the first term of (\ref{Zk}) is by far dominant for superhorizon scales, 
because \mbox{$k\ll 2H$}. Thus, for the superhorizon spectrum of $Z$ we have,

\begin{equation}
|Z(k)|
%=\frac{H}{2\pi}\Big(\frac{M_Z}{H}\Big)^2
%\,\Big(\frac{k}{2H}\Big)^{(M_Z/H)^2}
\simeq\frac{H}{2\pi}\Big(\frac{M_Z}{H}\Big)^2
\end{equation}
i.e. the $Z$-boson has an {\em almost scale invariant superhorizon spectrum} 
(plus a logarithmic tilt). In contrast, the photon $A_\mu$ is not coupled to 
any scalar field and its mass is exactly zero. Therefore, if we set
$A$ in place of the $Z$ in (\ref{Zk}), then \mbox{$M_A=0$} gives 
that, \mbox{$A(k)=k/4\pi\ll Z(k)$}. Thus, on superhorizon scales the amplitude 
of the $Z$-boson is much larger that the conformal invariant photon field, as 
shown in Fig.~1.

\begin{figure}
\parbox{0in}{\leavevmode
\hbox{\epsfxsize=2.5in
\epsffile{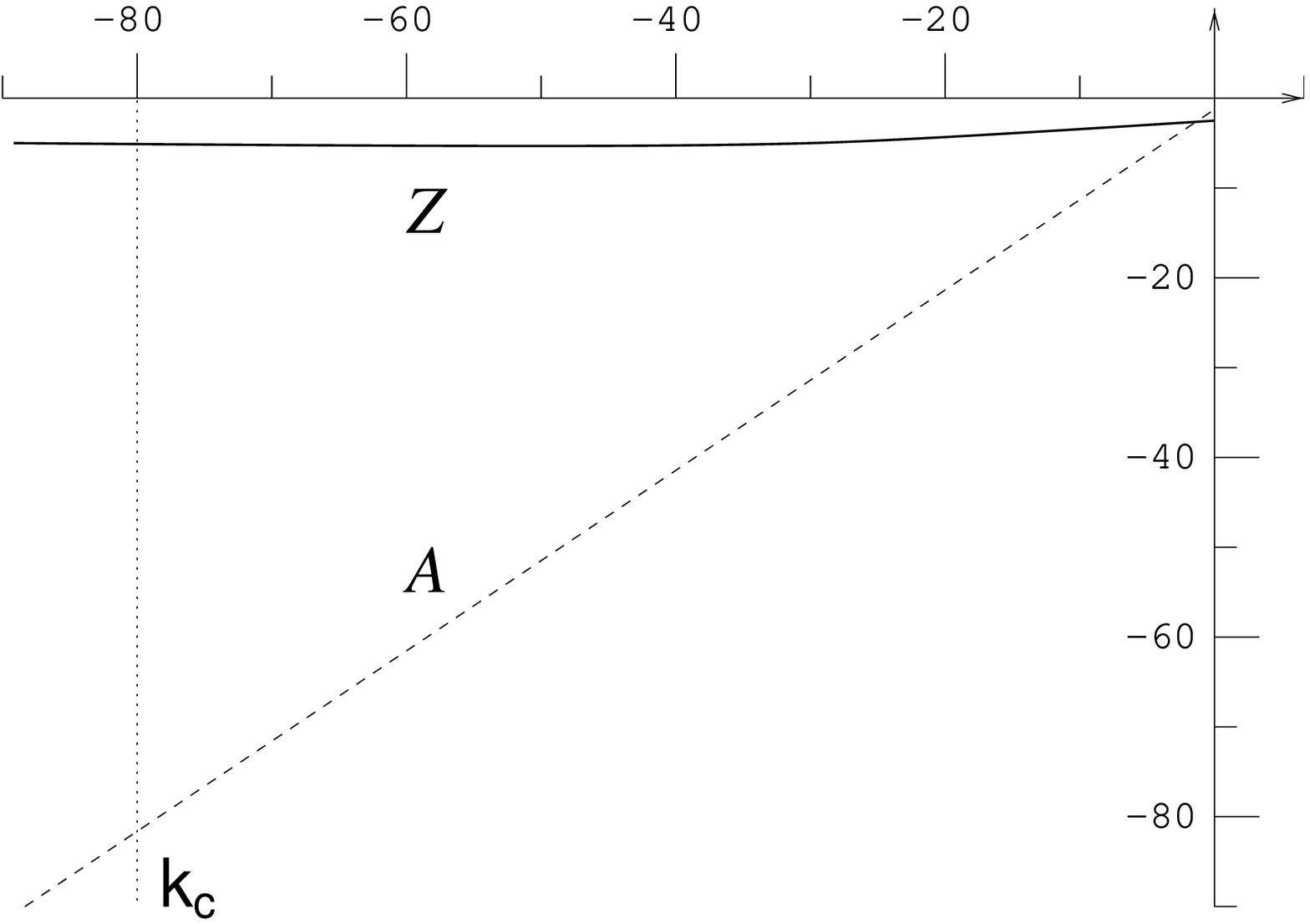}}
\begin{picture}(10,10)
\put(175,132){ln($k/H$)}
\put(110,35){\small ln[$Z(k)/H$]}
\put(110,20){\small ln[$A(k)/H$]}
\end{picture}}
\parbox{0in}{\narrowcaption{\footnotesize 
The superhorizon spectrum of the conformally invariant photon $A$ 
versus the one of the gravitationally generated $Z$-boson field. At the scale
of interest $k_c$ the difference is over 30 orders of magnitude.}}
\end{figure}

\section{The Magnetic Field at Galaxy Formation}

\paragraph{Evolution during the Hot Big Bang}

At the end of inflation reheating restores the EW-symmetry 
and the photon $A_\mu$ merges with the $Z_\mu$ to form the Hypercharge,

\begin{equation}
Y_\mu=\cos\theta_WA_\mu-\sin\theta_W Z_\mu
\end{equation}
where for the Weinberg angle of the SM,
\mbox{$\sin^2\theta_W\simeq 0.231$}. Now, since the photon production 
during inflation is negligible, {\em the Hypercharge spectrum is simply a 
projection of the $Z$-spectrum}, i.e. \mbox{$Y_\mu\simeq\sin\theta_W Z_\mu$}.

The Hypercharge is a massless, Abelian gauge field, which obeys the analog of 
Maxwell's equations. The associated hypermagnetic field is defined as 
\mbox{{\boldmath $B$}$^Y$$\equiv$ {\boldmath $\nabla$}$\times${\boldmath $Y$}},
so that, at the end of inflation we have, 

\begin{equation}
B^Y_{\rm rms}\simeq k(t_{\rm end})\,Y_{\rm rms}\simeq
k(t_{\rm end})\,\sin\theta_W\,Z_{\rm rms}
\label{BYend}
\end{equation}
Due to the high conductivity of the reheated plasma, {\em the hypermagnetic 
field gets frozen in and evolves satisfying flux conservation}, i.e. 
\mbox{$B^Y_\mu\propto a^{-2}$}. On the other hand, the relevant
hyperelectric component decays being Debye screened. Furthermore,
the magnetic field associated with the three $W_\mu^\alpha$ gauge fields of 
the SM, which are also gravitationally produced during 
inflation, is screened by the existence of a thermal mass, due to the 
self-coupling of the non-Abelian $W$-boson fields.

As the Universe continues to expand it cools down. 
When the temperature drops below the EW energy scale
the EW-symmetry is broken again and the photon is formed by
$Y_\mu$ and the non-Abelian $W^3_\mu$-boson,

\begin{equation}
A_\mu=\sin\theta_W W^3_\mu+\cos\theta_W Y_\mu
\end{equation}
Since the $W$-bosons are screened, their amplitude is negligible compared to 
the Hypercharge. Thus, the spectrum of the photon reflects that of the 
Hypercharge, i.e.\mbox{$A_\mu\simeq\cos\theta_W Y_\mu$}. Therefore, 
{\em at the EW-phase transition, the hypermagnetic field transforms into a 
regular magnetic field} as,

\begin{equation}
B_\mu=\cos\theta_WB^Y_\mu
\label{Breg}
\end{equation}
At first sight, it may strike as unlikely that a magnetic field is obtained
from the gravitational production of the $Z$-boson, which is orthogonal 
to the photon. In fact, the situation is analogous to that of light 
polarizers (Fig.~2). Light cannot go through two orthogonal polarizers.
However, when a third polarizer is inserted at an angle $\theta$, 
photons do cross. 

\begin{figure}
\parbox{0in}{\leavevmode
\hbox{\epsfxsize=1.8in
\epsffile{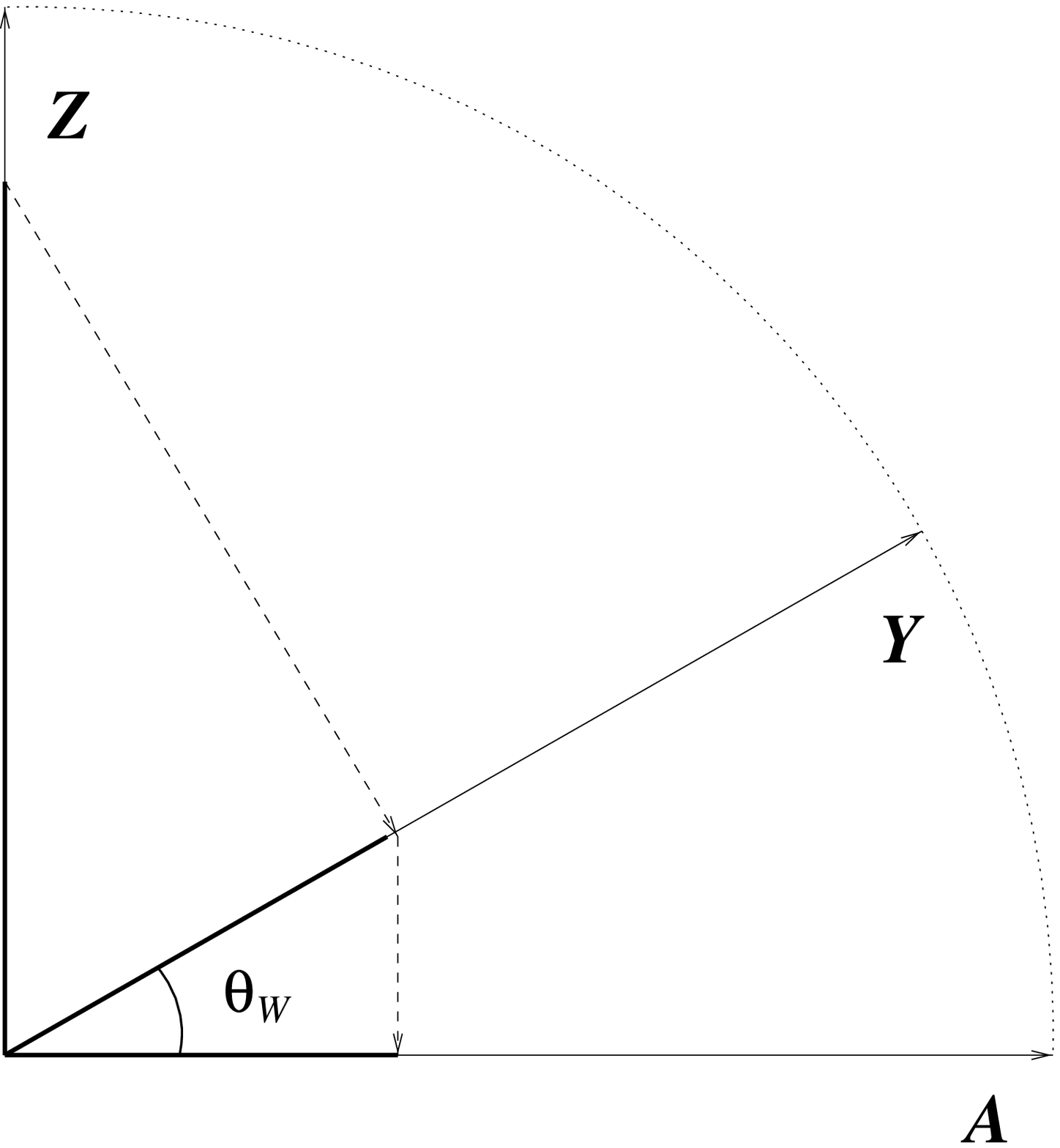}}}
\parbox{0in}{\narrowcaption{\footnotesize
The formation of a magnetic field from the generation of the $Z$-boson
has to pass through the intermediate Hypercharge stage similarly to 
light when crossing a set of orthogonal polarizers, which needs an 
intermediate third polarizer at some angle $\theta$ in order to go through.}}
%\vspace{-5mm}
\end{figure}

\paragraph{The magnitude of the Seed Field}

\mbox{$k(t_{\rm end})$} scaled until today is,

\begin{equation}
k(t_{\rm end})=\frac{2\pi}{\ell}
(\frac{T_{\rm reh}}{T_{\mbox{\scriptsize\sc cmb}}})
\label{k}
\end{equation}
where $\ell$ is the scale of the mode in question at present, 
$T_{\mbox{\scriptsize\sc cmb}}$ is the temperature of the CMBR at present,
\mbox{$T_{\rm reh}\simeq T(t_{\rm end})$} (prompt reheating) and we used that
\mbox{$a\propto T^{-1}$} at all times. Assuming that the field remains 
frozen until galaxy formation and using (\ref{BYend}), (\ref{Breg}) and 
(\ref{k}) we find,

\begin{equation}
B^{\rm gf}_{\rm rms}=\pi\sin(2\theta_W)(1+z_{\rm gf})^2
\frac{T_{\mbox{\scriptsize\sc cmb}}}{\ell}
\frac{Z_{\rm rms}}{T_{\rm reh}}
\label{B}
\end{equation}
where $z_{\rm gf}$ is the redshift that corresponds to galaxy formation. 
The collapse of matter into galaxies amplifies the above
by a factor given by the fraction of the galactic matter 
density today to the present critical density, 
\mbox{$(\rho_{\rm gal}/\rho_0)^{2/3}\approx 5\times10^3$}. In view of this
(\ref{B}) becomes,

\begin{equation}
B^{\rm gf}_{\rm rms}=7.3\times 10^{-27}\left(\frac{\mbox{1 Mpc}}{\ell}\right)
\frac{Z_{\rm rms}}{T_{\rm reh}}\,\mbox{Gauss}
\label{Bo}
\end{equation}
\nopagebreak[4]
where  \mbox{$T_{\mbox{\scriptsize\sc cmb}}=2.4\times 10^{-13}$ GeV},
\mbox{$z_{\rm gf}\simeq 4$} and \mbox{$\sin(2\theta_W)\approx 0.84$}. 

The superhorizon spectrum of $Z_{\rm rms}$ is approximately scale invariant 
with \mbox{$Z_{\rm rms}\sim g_{\rm z}^2H/(2\pi)^3$} and 
\mbox{$H=H(t_{\rm end})$}. 
In typical inflationary models \mbox{$T_{\rm reh}\!\sim\!10^{16}$GeV} and
\mbox{$H(t_{\rm end})\!\sim\!\sqrt{G}\,T_{\rm reh}^2\!\sim\!10^{13}$GeV}. 
Considering
that the scale of the largest turbulent eddy corresponds to the comoving
scale of \mbox{$\ell_c\simeq 10$ kpc} before the gravitational collapse 
of the protogalaxy, we find, 

\begin{equation}
B_{\rm seed}\sim 10^{-30}\mbox{Gauss}
\label{B1}
\end{equation}
This is sufficient to trigger the galactic dynamo in the case of a spatially 
flat, dark-energy dominated Universe (\cite{dlt}).
Extra amplification ($\sim 10^{3}$) may be achieved by preheating 
(\cite{ours}). Also, additional enhancement is possible, when considering 
turbulent helicity phenomena (\cite{helic1}, \cite{helic2}).
%The results are summarized in Fig. 3.

%\begin{figure}[t!]
%\begin{center}
%\leavevmode
%\hbox{\epsfxsize=4in\epsffile{results.eps}}
%\caption{\footnotesize 
%Magnetic field spectra and relevant seed field bounds. In dash-dot-dot-dot
%(green) 
%is the vacuum spectrum \mbox{$B_\ell\propto\ell^{-2}$} obtained from 
%preheating, with amplification factor $\sim\!10^5$. 
%At \mbox{$\ell_c=10$ kpc},
%\mbox{$B_{\ell_c}\sim 10^{-50}$Gauss}. 
%In dots and solid 
%(red) 
%is the spectrum \mbox{$B_\ell\propto\ell^{-1}$} from inflation in our 
%mechanism, {\em with} and {\em without} preheating amplification. 
%For this spectrum, \mbox{$B_{\ell_c}\sim 10^{-29}$Gauss} and
%\mbox{$B_{\ell_c}\sim 10^{-34}$Gauss} respectively. 
%Also, in dash-dots
%(blue)
%is the spectrum enhanced by helical turbulence (at $\ell_c$ the enhancement 
%factor is $\sim$20). These are to be compared with the dynamo bounds (rescaled 
%by a factor $5\times 10^3$) \mbox{$B_{\rm seed}\sim 10^{-27}$Gauss} and
%\mbox{$B_{\rm seed}\sim 10^{-34}$Gauss} corresponding to a spatially-flat,
%critical matter-density and dark-energy dominated Universe 
%respectively.}
%\end{center}
%\vspace{-1cm}
%\end{figure}

\section{Conclusions}

We have shown that all inflationary models of grand unification scale 
create magnetic fields of enough strength and coherence to trigger 
successfully the galactic dynamo.
Since this is a {\em model-independent magnetogenesis mechanism} it can 
be thought of as a feature of inflationary theory itself. Thus,
{\em accounting for the observed galactic magnetic 
fields can be considered as another generic success of inflation}.

\begin{acknowledgments}
\noindent
Work done in collaboration with A.C.Davis, T.Prokopec
and O.T\"{o}rnkvist. 
\end{acknowledgments}

\bibliographystyle{apalike}

\end{document}